\documentclass{aa}

\usepackage{graphicx}
\usepackage{amsmath,amssymb}    
\usepackage[varg]{txfonts}
\usepackage{natbib,twoopt}
\usepackage{upgreek}
\usepackage[colorlinks=true, allcolors=blue]{hyperref}
\usepackage{orcidlink}
\let\orcid\orcidlink

\bibpunct{(}{)}{;}{a}{}{,}

\makeatletter
\renewcommand*\aa@pageof{, page \thepage{} of \pageref*{LastPage}}
\makeatother

\begin{document}

   \title{Eye of the beholder: Observer reference frame bias in\\ 
   Hickson-like compact groups of galaxies}


   \author{Ariel Zandivarez\orcid{0000-0003-1022-1547}\inst{1,2}\fnmsep\thanks{\email{ariel.zandivarez@unc.edu.ar}}
          \and
          Eugenia Díaz-Giménez\orcid{0000-0001-5374-4810}\inst{1,2}
          \and
          Ailen R. Callen\orcid{0009-0006-6007-3115}\inst{1,2}
          }

   \institute{Universidad Nacional de C\'ordoba (UNC). Observatorio Astron\'omico de C\'ordoba (OAC). C\'ordoba, Argentina
         \and
             CONICET. Instituto de Astronom\'ia Te\'orica y Experimental (IATE). Laprida 854, X5000BGR, C\'ordoba, Argentina
             }

   \date{Received XXX XX 2025 / Accepted XXX XX 2026}

 
  \abstract
   {The identification of compact groups (CGs) is a recurring topic. Hickson's observational criteria are among the most widely used to identify CGs, yet the low membership in these systems suggests that these constraints may not hold when different observers in redshift space attempt to identify them.
   }
   {We investigate how the identification of Hickson-like CGs depends on the observer’s reference frame, quantifying how frequently the same system would be recognised from different vantage points.
   }
   {Using a mock lightcone built from the Millennium I Simulation plus a semi-analytic model of galaxy formation, we identified 7709 CGs when applying the standard Hickson-like criteria. For each group, we placed 1000 random observers on a surrounding sphere, recomputed redshift space coordinates and magnitudes, and reapplied the velocity and compactness requirements to test recoverability. We also examined the variation of population and local isolation against different observer directions.  
   }
   {The velocity concordance criterion shows modest sensitivity to the observer's location: About 10\% of CGs fail for some observers, typically groups with members with high peculiar velocities ($>1000 \ {\rm km \ s^{-1}}$). The compactness requirement is far more fragile, as $\sim 44\%$ of CGs are missed by most observers, and these systems are very elongated or are chance alignments in real space. Tightening selection limits reduces this dependence. Lowering the surface brightness threshold to $\mu \leq 23 \ {\rm mag \ arcsec^{-2}}$ reduces the compactness dependence to $\sim 16\%$, while reducing the velocity limit to $\Delta V\leq250 \ {\rm km \ s^{-1}}$ lowers velocity-driven failures to less than 4\%. Applying both cuts simultaneously yields up to 84\% observer-independent groups, although with a substantially smaller sample. 
   Population and isolation are affected by bright interlopers seen from different directions. While such interlopers are common, they have only a minor effect on the compactness and velocity concordance criteria; however, the local isolation is commonly broken.
    }
   {Observer frame effects, dominated by the compactness criterion, can significantly bias Hickson-like CG catalogues. However, adjusting surface brightness and velocity difference thresholds allows users to balance the physical reliability according to their specific scientific goals.
   }

                
   \titlerunning{Observer frame bias in CGs}
   \authorrunning{A. Zandivarez, E. D\'iaz-Gim\'enez \& A.R. Callen }
   \maketitle
%

\section{Introduction}
Compact groups (CGs) are relatively isolated associations of a small number of galaxies in close proximity, and they provide an ideal setting for the investigation of galaxy interactions. The most widely known catalogue of such systems was compiled by \cite{Hickson82}, who visually identified CGs on the plane of the sky as associations of at least four bright galaxies with high mean surface brightness and no nearby bright companions. A decade later, when redshift information became available, \cite{Hickson92} refined the sample to reduce contamination from interlopers. This overall selection method is commonly known as the Hickson criteria.

Several works have identified CGs based on Hickson's recipes (e.g. \citealt{Prandoni+94,Iovino+02,Lee+04,McConnachie+08,McConnachie+09,DiazGimenez&Mamon10,DiazGimenez+12,sohn+15,DiazGimenez+18,zheng+20,zandivarez+24}). Among the different studies that have been performed on CGs, we can mention those related to their environment (e.g. \citealt{mendel+11,DiazGimenez+15,zheng+21,taverna+23}), luminosity functions (e.g.  \citealt{zepf97,hunsberger98,coenda+12,zheng+21,zandivarez+22}), gravitational lensing \citep{chalela+17,chalela+18}, suppression of star formation (e.g. \citealt{Bitsakis+16,lisenfeld+17}), and different galaxy populations, morphologies, and stellar content (e.g. \citealt{zucker+16,moura+20}). On the other hand, this procedure has also been applied to identify CGs in synthetic galaxy catalogues on several occasions. In recent years, using CGs identified in mock catalogues, a series of detailed investigations have been performed that have included the analysis of their frequency and nature \citep{DiazGimenez+20}, their assembly channels \citep{DiazGimenez+21}, the performance of the Hickson-like automatic algorithm utilised in flux-limited catalogues \citep{Taverna+22}, the time evolution of the properties of galaxy members based on their assembly channels \citep{zandivarez+23}, the location of CGs characterised by specific assembly channels within the large-scale structure of the Universe \citep{taverna+24}, and the evolution of their two most massive galaxies \citep{zandivarez+25}.

Beyond their widespread use in extragalactic studies, it has long been recognised that the Hickson identification criteria have limitations. Originally, they were designed to select small, physically dense systems of closely spaced galaxies that are relatively isolated in the Universe. Over time, however, it has been shown that a significant fraction of the resulting groups are not physically dense systems (e.g. \citealt{McConnachie+08,DiazGimenez&Mamon10}). Following the classification of CGs presented in \cite{DiazGimenez&Mamon10} based on real 3D sizes, \cite{DiazGimenez+20} demonstrated, using several semi-analytic models of galaxy formation, that only about 35\% to 55\% of the groups in a catalogue are expected to be physically dense in real space, while the remainder are likely chance alignments in the line of sight. This issue was noted early on by \cite{hickson+97}, who pointed out that a sample of systems selected by surface density is subject to different biases. Geometric bias arises because non-spherical systems are more likely to be chosen when orientated to present a smaller projected area along the line of sight. Kinematic bias is due to a high-density temporary phase of the system caused by the orbital motions of its members. The combined effect causes either a chance alignment or a transiently compact configuration within a looser association \citep{Rose77,rose79,Mamon86}. 

In particular, when focusing on the identification process following the Hickson criteria, the two conditions most closely tied to system geometry are those of compactness (surface brightness) and velocity concordance (velocity difference with respect to the centre). These criteria depend directly on what the observer can measure from their reference frame in redshift space. 
The compactness requirement is clearly affected by the geometric bias noted by \cite{hickson+97} since an observer may detect a system that is intrinsically elongated in real space but happens to present a minimal cross-section along their line of sight. 
Likewise, working in redshift space means that galaxy positions along the radial direction are distorted by the projection of their peculiar velocity vectors onto the observer’s line of sight. As a result, an observer might conclude that the galaxies are genuinely close together or identify the system because the peculiar velocity vectors have little radial component. 
This can occur either because the galaxies have low peculiar velocities or because their velocity vectors are considerably aligned with the plane of the sky. Consequently, the observer’s location relative to the system can have a substantial impact on its detection. Moreover, given the very low membership of these groups, typically only three or four galaxies, the position or the peculiar velocity vector of a single member can determine whether a particular observer detects the system.

On the other hand, changing the observer’s position can also affect other components of the Hickson criteria, namely the population and isolation criteria. These criteria are not directly related to the compact nature of the system, but they impose minimum and maximum membership requirements and demand local isolation from nearby bright galaxies. When the observer’s position changes, additional bright galaxies may appear along the new line of sight in the redshift space that did not interfere with the identification of the CG along the original viewing direction. These galaxies may now fall within the system itself or lie outside it but within the region where isolation from bright galaxies is required, thereby affecting the fulfilment of both criteria.

Consequently, questions arise regarding what fraction of CGs identified with the Hickson criteria are truly independent of the observer’s location in the Universe and what physical conditions characterise the systems that remain observable from any vantage point. Therefore, in this work, we aim to quantify the influence of the observer’s position on the identification of CGs selected according to the Hickson criteria using a catalogue of synthetic CGs identified on a mock lightcone.  

The layout of this work is as follows. In Sect.~\ref{sec:samples}, we present the mock catalogue used in this work and the sample of CGs. In Sect.~\ref{sec:results}, we describe the experiment to quantify the relevance of changing the observer's vantage point to recover the original sample of the CGs. We summarise and discuss our results in Sect.~\ref{sec:conclusions}.

\section{Samples}
\label{sec:samples}
We used a sample of CGs built in previous works (\citealt{DiazGimenez+20} and other papers of that series). Here, we briefly describe the simulation and methods used in this work. 

\subsection{Mock catalogues}
The mock catalogue is a lightcone built from the Millennium I Simulation \citep{Springel+05}. The dark matter cosmological simulation is combined with a semi-analytic model of galaxy formation to assign synthetic galaxies.\footnote{\url{http://gavo.mpa-garching.mpg.de/Millennium/}} The semi-analytic model used in this work is the one built by \cite{Guo+11} applied on top of a simulation box with side of $500 \ h^{-1} \, {\rm Mpc} $ and a WMAP1 cosmology \citep{Spergel+03}. This model broadly reproduces several basic, observed galaxy properties and statistics. 

We use this sample to build a mock galaxy lightcone following \cite{jpas}. We mimic the evolution of properties and structures in the Universe by stacking snapshot slices of a given width representing different cosmic epochs. We place an observer, and the space is distorted along its radial direction, considering the expansion of the Universe and the peculiar velocities of each object. We use a K-decorrection procedure \citep{DiazGimenez+18} to estimate the observer-frame apparent magnitudes from the rest-frame absolute magnitudes provided by the simulation. The final mock lightcone is an all-sky survey with $z\leq0.2$ and an observer-frame apparent SDSS (Sloan Digital Sky Survey) r-band magnitude $r\leq17.77$. Following previous works, only galaxies with stellar masses larger than $\sim 10^9 \mathcal{M}_\odot$ were included.  

\subsection{The Hickson-like CG sample}
\label{sec:cgsid}
The identification of CGs is performed using a Hickson-like finding algorithm that searches for galaxy systems that satisfy constraints on membership, compactness, relative isolation, velocity concordance, and flux limit of the brightest group galaxy (BGG) \citep{zandivarez+22}.
The identification criteria stipulate the following: 
\begin{itemize}
    \item Population: There are at least three galaxies and a maximum of ten (including the brightest) within a three-magnitude range from the brightest.
    \item Flux limit of the BGG: To maximise completeness \citep{Prandoni+94,DiazGimenez&Mamon10}, the BGG of the system has to be at least three magnitudes brighter than the catalogue magnitude limit ($r_{\rm lim}=17.77$).
    \item Velocity concordance: Galaxy members are within $1\,000 \rm \, km\,s^{-1}$ from the median velocity of the system \citep{Hickson92}. 
    \item Compactness: The group surface brightness in the $r$-band is less than $26.33 \, \rm mag \, arcsec^{-2}$. 
    \item Relative or local isolation: There are no other bright galaxies (in a three-magnitude range from the brightest, nor brighter) within three times the radius of the smallest circle that encloses the centres of all galaxy members.
\end{itemize}

Since our galaxies are sizeless particles, we also included a recipe to deal with the possible observational blending of galaxies. We used the galaxy stellar mass estimated by the semi-analytic model to assign half-light radius \citep{Lange+15} to each dimensionless mock galaxy and consider two mock galaxies as blended if their radii overlap in projection. Finally, the total number of CGs identified in the lightcone is 7709 systems. 

\begin{figure}
   \centering
   \includegraphics[width=0.40\textwidth]{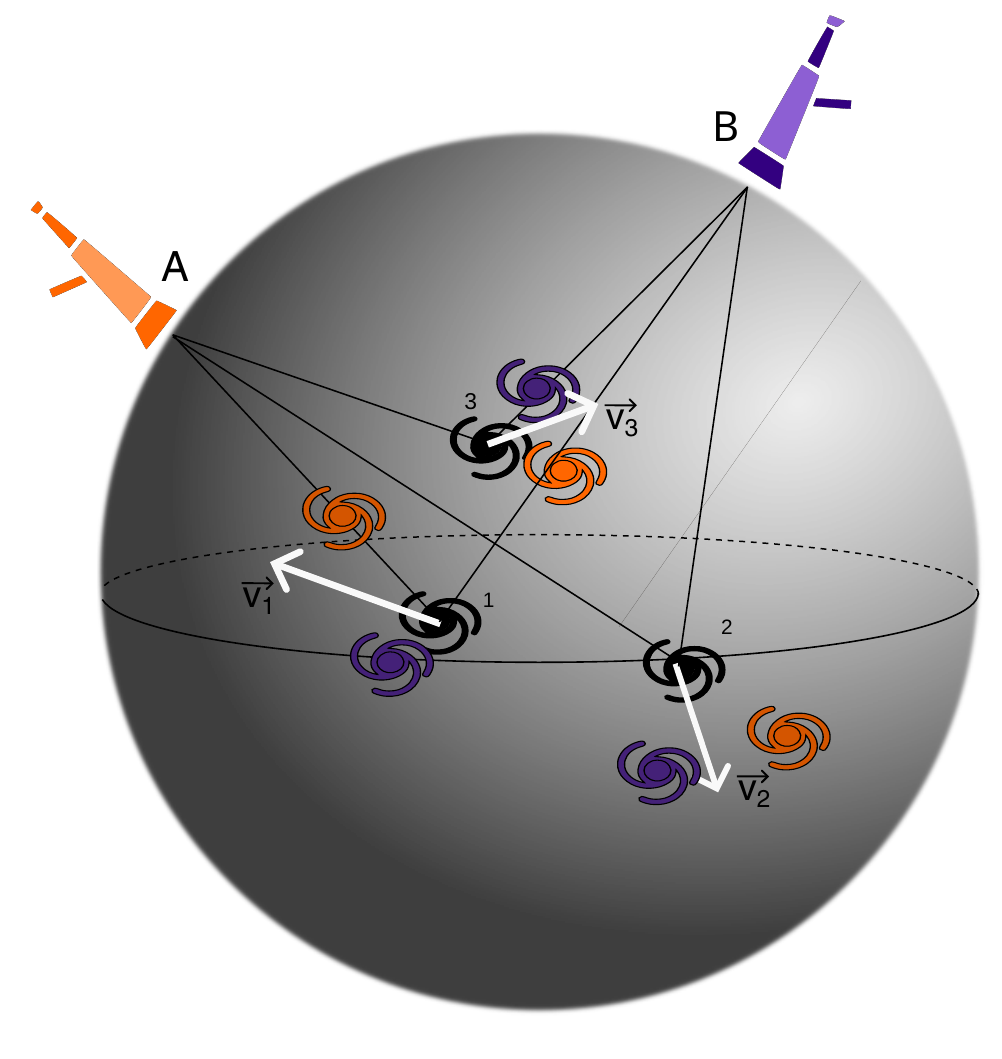}
      \caption{Illustration showing two observers viewing a CG from different directions. The black galaxies represent the triplet in real space, and the white arrows indicate their respective peculiar velocities. Each observer sees the galaxies in redshift space with positions distorted according to the projection of their peculiar velocity vectors along the line of sight (galaxies in colours). 
      Observer A (orange) views the triplet from a direction where the projected area on the sky is smaller, but the system is elongated along the line of sight, whereas Observer B (violet) perceives it as more extended in projection and not as elongated along the radial direction.}
         \label{fig:1}
\end{figure}

\section{How likely is it that different observers will identify the same Hickson-like CG?}
\label{sec:results}
As stated previously, we are interested in assessing the relevance of the observational rest-frame at which the observer is located. Given the very low membership of CGs (mostly three or four galaxy members) and the intrinsic collapse of the velocity information in the line of sight in redshift space, it is expected that a particular location of the observer could be relevant in the identification of a Hickson-like CG that fits all the set of restrictions. Hence, our main goal is to quantify whether a given CG identified with a Hickson-like algorithm at the rest-frame observer system could also be identified from a different rest-frame system at a different location in space. 

\subsection{Randomising the observer locations}
\label{sec:randomexp}
To achieve this goal, we devised an experiment using the sample of CGs identified with a Hickson-like algorithm in a mock lightcone (see previous section). Since the sample of CGs is identified in a mock catalogue, we have access to all the real space information of each system, i.e. their 3D real space positions, their peculiar 3D velocity vectors, and their rest-frame photometry. 

The procedure to randomise the observer point of view is as follows. For each of the 7709 CGs identified in redshift space, we determined its centre coordinates in 3D real space. We also measured the 3D positions of each galaxy member relative to the group centre in real space. Then, we only modified the 3D centre location by randomly selecting their polar coordinates with a uniform distribution (right ascension in $[0,2\pi]$ and the sine of declination in $[-1,1]$), keeping their original comoving distance to the observer ($d_{\rm cm}$). We used $d_{\rm cm}$ and the new right ascension and declination to compute the new 3D centre of the system. Then, for each galaxy, we added to this new 3D centre its relative positions to the original centre to determine its new 3D coordinates. Using these new 3D positions ($\vec{d}_{\rm new}$) and the original peculiar velocity vector of each galaxy ($\vec{v}$), we determined their new distorted coordinates in redshift space.\footnote{The observed redshift is estimated as $z_{\rm obs}=(1+z_{\rm cos})(1+\vec{v}.\hat{d}_{\rm new}/c)-1$, with $z_{\rm cos}$ the cosmological redshift obtained from 3D coordinates in real space and $c$ the velocity of light in the vacuum \citep{harrison74}.} Observer-frame apparent magnitudes were computed using the new redshifts of each galaxy. Finally, we checked whether this set of galaxies seen from a different location meets the velocity concordance and compactness criteria with which the original CG was identified.

This procedure was performed 1000 times for each CG originally identified in redshift space. 
The overall effect is the same as randomly placing 1000 observers around each CG on the surface of a sphere of radius equal to the comoving distance to the CG centre. This is schematised in Fig.~\ref{fig:1}. Then, for each new observer, the velocity and compactness criteria were tested according to the observable properties seen from their own line of sight. Groups that no longer meet any of those criteria were considered missing from the perspective of that observer.  
It is important to emphasise that whenever the velocity concordance and compactness criteria were required, they were applied exactly as defined in Section~\ref{sec:cgsid}. These limits, established during the identification process, are the ones that are enforced consistently throughout this study.

\begin{figure}
   \centering
   \includegraphics[width=0.49\textwidth]{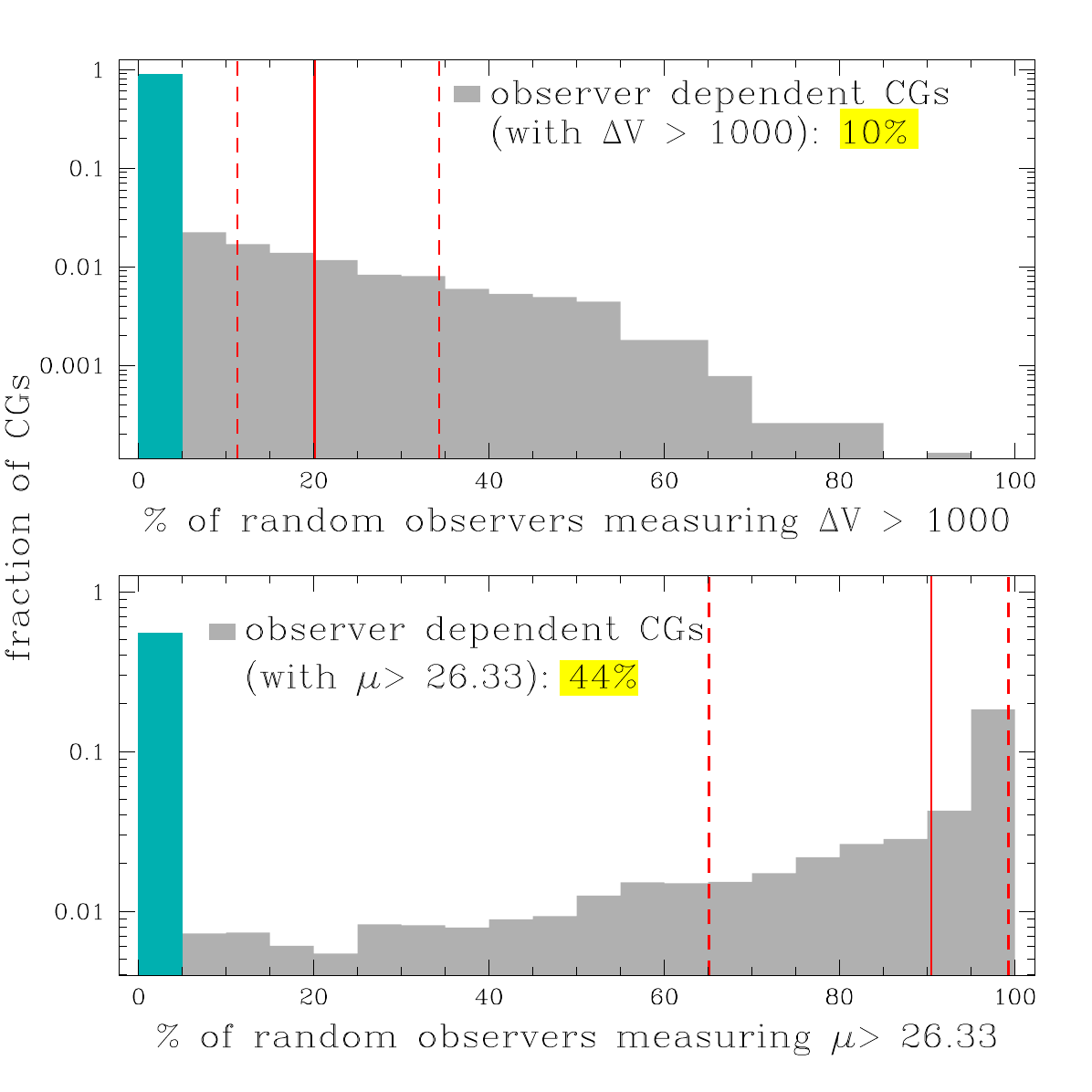}
      \caption{Distribution among the 7709 CGs of the percentage of 1000 virtual observers seeing the CG fail the selection criteria.
      Top panel: Percentage of all-sky observers who lose the CGs because of the radial velocity filter. Bottom panel: Same but when analysing the compactness criterion (surface brightness, $\mu$). The inset yellow legends in both panels quote the percentage of the total sample of CGs that are lost for at least 5\% of randomly orientated observers. Vertical red lines represent the medians (solid) and the 25th and 75th percentiles (dashed) of the distributions for the observer-dependent CGs (grey histograms). The teal bar between 0 and 5\% represents the fraction of CGs that are considered robust.}
         \label{fig:2}
\end{figure}

\subsection{How many CGs would meet the velocity concordance or compactness criteria?}
In Fig.~\ref{fig:2}, we show the fraction of CGs as a function of the percentage of random observers for which at least one of the Hickson criteria is no longer satisfied.   
In the top panel, only the velocity concordance is evaluated, while the compactness criterion is considered alone in the bottom panel. 
We set a maximum tolerance limit to determine when a given CG can be considered independent of the observer’s point of view. Specifically, we define a CG as observer-independent (or robust) when it fails any of the criteria in fewer than 5\% of all possible random observer positions (teal bar in the figure).
Therefore, the top panel quotes that the percentage of observer-dependent CGs is $\sim 10\%$, i.e. these are the CGs originally identified that will have problems being recovered by other observers when the velocity concordance criterion is required. 
The histogram shows the distribution of the percentages of random observers who have lost their CGs. The percentages for the observer-dependent CGs (grey histogram) range from very few to $\sim 85\%$ of random observers, with a median of $20\%$. 
In the bottom panel of Fig.~\ref{fig:2}, we quote that $\sim 44\%$ of the CGs are expected to be missed when the compactness criterion is tested from the perspective of the random observers.  
The percentage of random observers that could not identify the original CGs (histogram) has a median of $\sim 90\%$. Hence, the compactness criterion is very sensitive to the location of the observer; not only could almost half of the original CG sample be missed, but in those cases, almost all the observers will miss them. 

\begin{figure}
   \centering
   \includegraphics[width=0.49\textwidth]{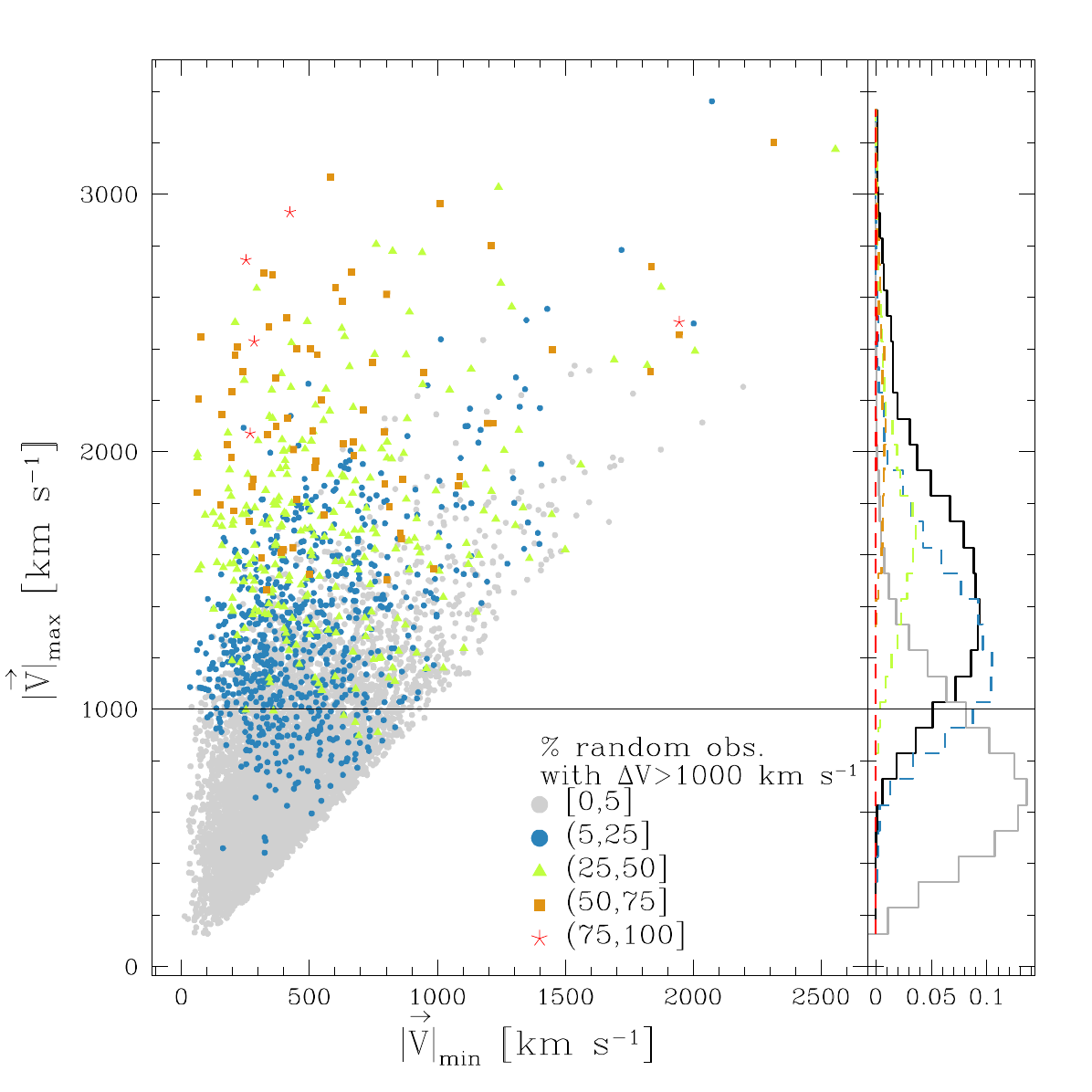}
      \caption{Scatter plot of the maximum and minimum galaxy peculiar velocity moduli for each CG. The grey dots represent the sample of CGs that are lost by fewer than 5\% of the observers when applying the velocity filter, while the coloured dots are the sample of CGs that are lost for larger percentages of random observers. Different colours represent different percentages of observers who lose each CG (inset legends). The right panel displays the histograms of the maximum velocity moduli for non-lost CGs in grey and lost CGs in black (the coloured histograms are sub-samples of the black).}
         \label{fig:3}
\end{figure}

We examined some of the system properties to investigate why changes in the observer’s location affect the identification of CGs with respect to the velocity and compactness criteria.
First, we analysed the peculiar velocities of the CG member galaxies. 
For each galaxy, we calculated the modulus of its peculiar velocity vector. For each system, we identified the members with the largest and smallest velocity modulus as a way of characterising the group. Figure~\ref{fig:3} shows the scatter plot of these maximum and minimum velocity moduli as a function of the percentage of random observers who can identify the CGs or not. 
Grey points correspond to the distribution of velocity moduli for CGs identified by almost all possible random observers, while the other colours represent CGs that some observers could not identify. 
The colour scale encodes the percentage of observers who failed to identify each system (see inset legends). The right panel displays the marginal distributions of the maximum velocity modulus for each system: the grey distribution corresponds to CGs identified by almost all the observers (at least 95\% of them), while the black distribution corresponds to those missed by more than 5\% of the 1000 observers. As expected, the figure shows that CGs identified by most of the random observers are concentrated at low peculiar velocity modulus values, whereas systems lost by more than 5\% of the observers tend to cluster at higher values. Using 1000 ${\rm km \ s^{-1}}$ as a reference for the maximum velocity modulus (black horizontal line), we find that 78\% of robust systems have velocities below this threshold, compared to only 10\% of observer-dependent systems. Thus, it is reasonable to conclude that CGs containing members with velocities above 1000 ${\rm km \ s^{-1}}$ are highly likely to have an identification strongly dependent on the observer’s location relative to the velocity vector.

Secondly, to analyse the large number of CGs that cannot be recovered by some observers when the compactness criterion is applied, we follow the approach proposed by \cite{DiazGimenez&Mamon10}. In that work, the authors analysed a sample of CGs identified with  Hickson-like criteria and examined how their real space sizes varied. This allowed them to define two main classes of CGs: physically dense systems (`Reals') and chance alignments along the line of sight (CAs). 
Their classification was based on two real space size measurements for the four closest members: the 3D maximum intergalaxy separation, $s_{\rm max}$, and the elongation computed as the ratio between the line-of-sight size and the projected size, $(s_{\parallel}/s_{\perp})$. They considered as Reals those CGs with $s_{\rm max} \le 100 \ h^{-1} \, {\rm kpc} $, or with $s_{\rm max}$ between 100 and 200 $ h^{-1} \, {\rm kpc}$ and $(s_{\parallel}/s_{\perp}) \leq 2$. The remaining systems were classified as CAs. Later, \cite{DiazGimenez+21} introduced a subdivision within the CAs, identifying systems that are very extended in real space and labelling them as `Fake'. These were defined as systems with $s_{\rm max} \ge1000 \ h^{-1} \, {\rm kpc} $ or with the minimum 3D intergalaxy separation larger than $200 \ h^{-1} \, {\rm kpc} $. In this work, we use these real space measures, but we extend them to include triplets as well. The percentage of Real systems for the total sample of CGs is $49\%$, while $37\%$ are classified as CAs and $14\%$ as Fake. 

\begin{figure}
   \centering
   \includegraphics[width=0.49\textwidth]{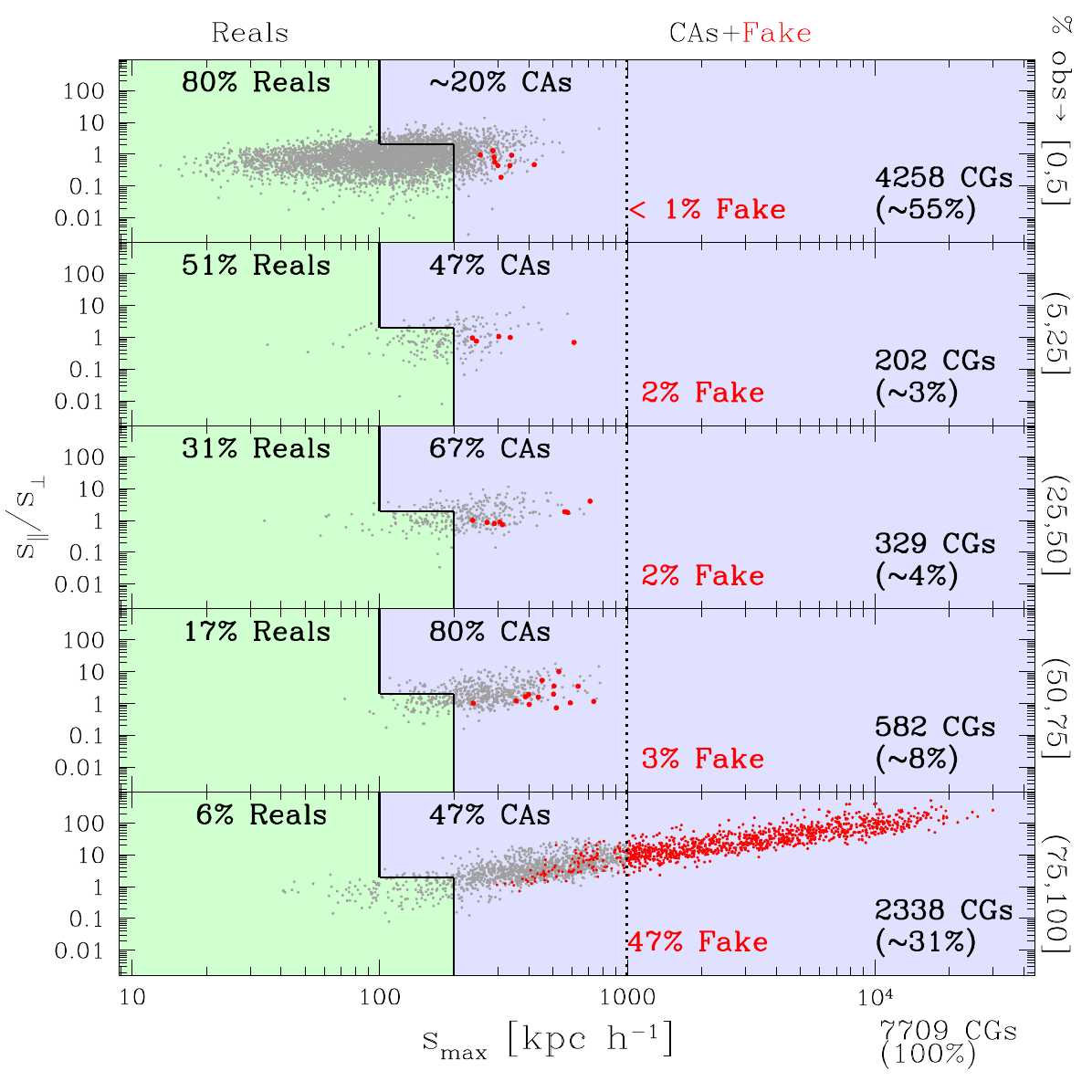}
      \caption{Scatter plot of the largest 3D real space separation in the CG ($s_{\rm max}$) for the three (in the case of triplets) or four (for the remaining CGs) closest galaxies and the original line-of-sight elongation, features used to classify into Reals and CAs \citep{DiazGimenez&Mamon10}. The light green region corresponds to Real CGs, while outside that region, CGs are classified as chance alignments (CAs). Within the CA class, we have also classified CGs as `Fake' if the maximum 3D separation is larger than $1 \  h^{-1} \, {\rm Mpc}$ (dotted vertical line) or the minimum 3D separation is larger than 200 $ h^{-1} \, {\rm kpc}$ (red dots). The top panel shows the distribution of CGs that satisfy the compactness criterion, almost regardless of the position of the observer (for more than 95\% of them). The remaining panels display the samples where some observers (more than 5\%) lose the CG because they no longer achieve the compactness criterion from their points of view. From the second row to the bottom, the panels display the CGs when the percentage of random observers who lose the CG is between 5 and 25; 25 and 50; 50 and 75; and 75 and 100.}
         \label{fig:4}
\end{figure}

In Fig.~\ref{fig:4}, we show the scatter plot of these sizes as a function of the percentage of observers who identify the systems when they require the compactness criterion. 
We outline the regions corresponding to CGs classified as Real (light green) and CAs+Fakes (lavender). The top panel shows only those CGs that were identified by almost all random observers when applying the compactness criterion, while the lower panels correspond to systems that were missed by more than 5\%  of the observers (the fraction of missed detections increases downward).

\begin{figure}
   \centering
   \includegraphics[width=0.49\textwidth]{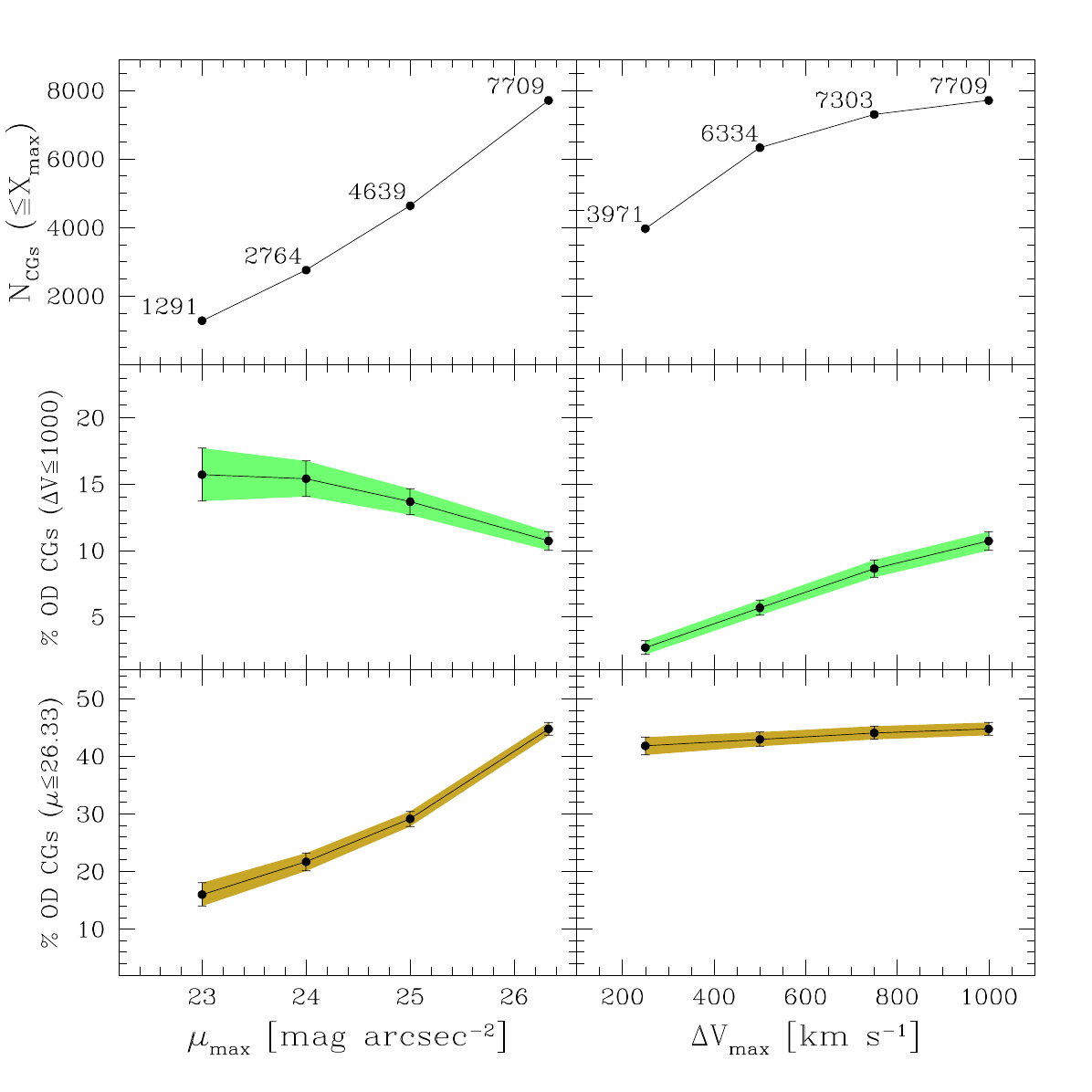}
      \caption{Percentage of observer-dependent (OD) CGs for four different sub-samples defined by varying the maximum surface brightness or the velocity gap in the line of sight. The left column shows the number of CGs (top panel), the percentages of CGs that are lost due to the velocity filter (middle panel), and $\mu$ criterion (bottom panel) as a function of the original $\mu_{\rm max}$ of the CG, i.e. each bin only takes into account CGs with an original $\mu$ less than the corresponding bin value ($\mu \leq \mu_{\rm max}$). 
      The right column shows the same as the left column, but for samples selected as a function of the maximum velocity difference of galaxy members from the CG centre ($\Delta V \leq \Delta V_{\rm max} )$. In each case, the last bin corresponds to the original values of $\mu_{\rm lim}$ and $\Delta V_{\rm lim}$ of the CG sample.}
         \label{fig:5}
\end{figure}

This figure confirms that, as expected, the real space size of CGs identified with a Hickson-like algorithm correlates with the probability that these systems are not recovered by different observers. Among the systems identified by almost all the observers, 80\% are classified as Reals, whereas among those missed by almost all observers (75\% or more), 94\% are catalogued as CAs or Fakes. However, we also find Real systems that are not identified by some observers, as well as some CAs that accomplish the compactness criterion regardless of the observer’s location. It is worth mentioning that the very few Fake systems that meet the compactness criterion (top panel) are those that exceed the minimum real space separation, not their maximum one (no red points at abscissa greater than 1000 ${\rm km \ s^{-1}}$).
The fact that some groups classified as Real were missed by certain observers is mainly due to the presence of more than four members. In such cases, the four closest galaxies form a compact system in the $s_{\rm max}$-$(s_{\parallel}/s_{\perp})$ plane, but the additional members can make the group appear elongated along a particular direction. 
This elongation makes the angular diameter larger seen from some direction, and therefore the surface brightness fainter, reducing the likelihood that the system will be identified by all possible observers.
Nevertheless, the main conclusion is that an elongated group in real space, selected along a favourable end-on view, will usually look elongated in randomised space, hence with a lower surface brightness. In other words, systems intrinsically elongated in real space are the most likely to be missed when the observer’s position changes.

An interesting outcome of this analysis is that a large fraction of the systems failing to satisfy the compactness criterion for all possible observers are precisely those associated with elongated structures in real space, i.e. systems that are not expected to be physically compact. This naturally raises the question as to whether there is a way to reduce the presence of such observer-dependent systems in a sample of CGs already identified with a Hickson-like algorithm.

\subsection{Minimising the presence of observer-dependent Hickson-like CGs}
We examine whether a sample of CGs already identified with a Hickson-like algorithm can be adjusted to minimise the influence of the observer’s reference frame. Our aim is not to perform a new identification of CGs, but rather to use the sample identified with the original criteria and apply a new set of restrictions. This approach was chosen with the idea of accounting for all observational catalogues identified using Hickson-like criteria found in the literature, which could be improved if appropriate tuning is applied.

In this work, we applied a set of cuts to the original sample using the limiting surface brightness and the limiting velocity difference as variables. We created sub-samples for four values of surface brightness limit and four values of velocity-difference limit. 
The left column of Fig.~\ref{fig:5} shows the sub-samples obtained when adopting surface brightness limits of 23, 24, 25, and the value of the original sample, 26.33 ${\rm mag \ arcsec^{-2}}$, while keeping the velocity-difference limit fixed at 1000 ${\rm km \ s^{-1}}$. 
The right column displays the results when varying the velocity-difference limit to 250, 500, and 750, as well as the original 1000 ${\rm km \ s^{-1}}$, while keeping the surface brightness limit fixed at 26.33 ${\rm mag \ arcsec^{-2}}$. The first row displays the number of CGs in each sub-sample. The second and third rows show the percentage of CGs that are observer-dependent under either the velocity-concordance criterion or the compactness criterion, respectively, as a function of the property being modified.

\begin{figure}
   \centering
   \includegraphics[width=0.49\textwidth]{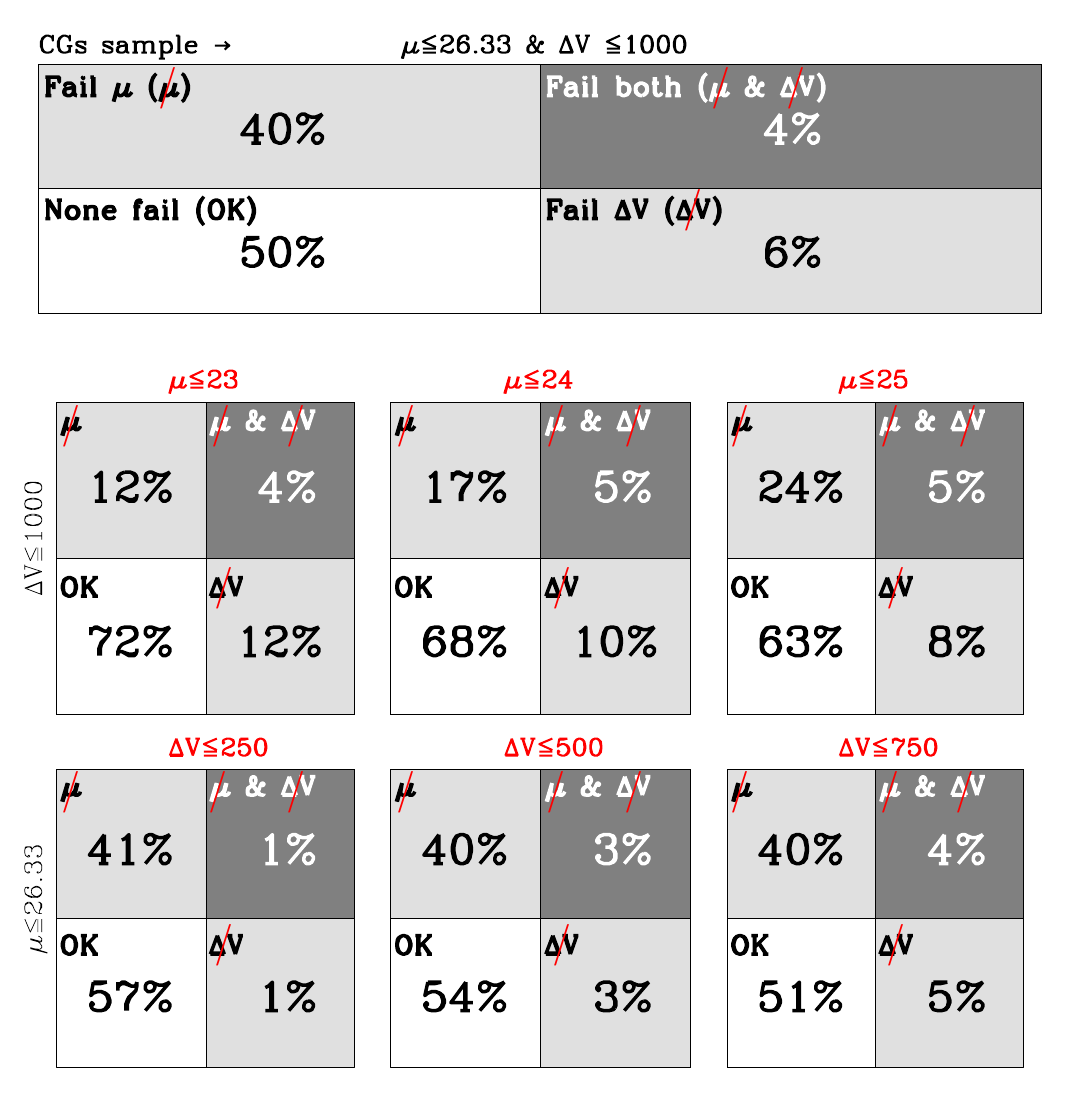}
      \caption{Percentages of CGs obtained when analysing the joint effect of whether an observer may or may not miss a CG due to surface brightness, velocity gap, or both. Dark grey areas highlight the percentage of CGs that random observers have lost due to the failure of both criteria, while light grey areas display the percentages obtained when the CGs are lost by exceeding the limits in surface brightness (26.33) or velocity gap (1000). Finally, white areas quote the percentage of CGs that are identified by at least 95\% of the random observers, regardless of their point of view. The top panel displays the percentages for the full sample of CGs, while the middle and bottom panels quote the percentages when different sub-samples are defined as a function of the surface brightness or the velocity gap, respectively.
      } 
         \label{fig:6}
\end{figure}

Defining sub-samples with higher surface brightness reduces the fraction of observer-dependent CGs for the compactness criterion (first column, third row), decreasing from the original 45\% to approximately 16\% for the sub-sample with a limit of $\mu=23$. 
However, this modification does not significantly improve the fraction of observer-dependent CGs for the velocity-concordance criterion, which varies from 11\% to 15\%. 
In contrast, varying only the velocity-difference limit lowers the fraction of observer-dependent CGs for the velocity-concordance criterion from the original 11\% to below 4\% when a limit of 250 ${\rm km \ s^{-1}}$ is adopted. Yet, changing this parameter produces no significant improvement for the fraction of observer-dependent CGs under the compactness criterion, which remains nearly constant at about 44\% on average. Therefore, the main result is that the relative decrease of the fraction of observer-dependent CGs is roughly a factor of three for both $\mu_{\rm lim}$ = 26.33 $\rightarrow$ 23 and $\Delta V_{\rm lim}$ = 1000 $\rightarrow$ 250 $\rm km \ s^{-1}$.

Up to this point, our analysis has shown how the percentages of observer-dependent CGs vary when the surface brightness and the velocity-difference maxima are changed, considering the compactness and velocity-concordance criteria separately. 
One may then ask what percentages are obtained when both criteria are required to be satisfied simultaneously, or in which cases each criterion fails. To address this, we used the same sub-samples defined above and examined the percentages when both criteria are applied together.
Figure~\ref{fig:6} presents these results. The lower-left quadrant shows the percentage of CGs that satisfy both criteria, the upper-right quadrant shows the percentage for which both criteria fail, depending on the observer, and the remaining quadrants indicate cases where only one criterion fails: compactness (upper left) or velocity concordance (lower right).

The top panel shows the result for the full CG sample. Roughly 50\% of the systems can be identified by any observer regardless of position. The dominant effect comes from failure of the compactness criterion, with 40\% of the systems depending on the observer. Only 4\% are missed because both criteria fail simultaneously, and another 6\% are lost solely due to the velocity-concordance criterion.

The middle and lower panels show the same scheme for the CG sub-samples obtained by varying the surface brightness (middle panels) or the velocity-difference maximum (lower panels). 
These results indicate that restricting the surface brightness of the original CG increases the fraction of systems satisfying both criteria simultaneously, mainly at the expense of a reduction in the fraction of systems lost due to failure of the compactness criterion. A similar, though less pronounced, improvement is seen when lowering only the velocity-difference limit, which primarily reduces the fraction of systems that fail the velocity-concordance criterion.

\begin{figure}
   \centering
   \includegraphics[width=0.48\textwidth]{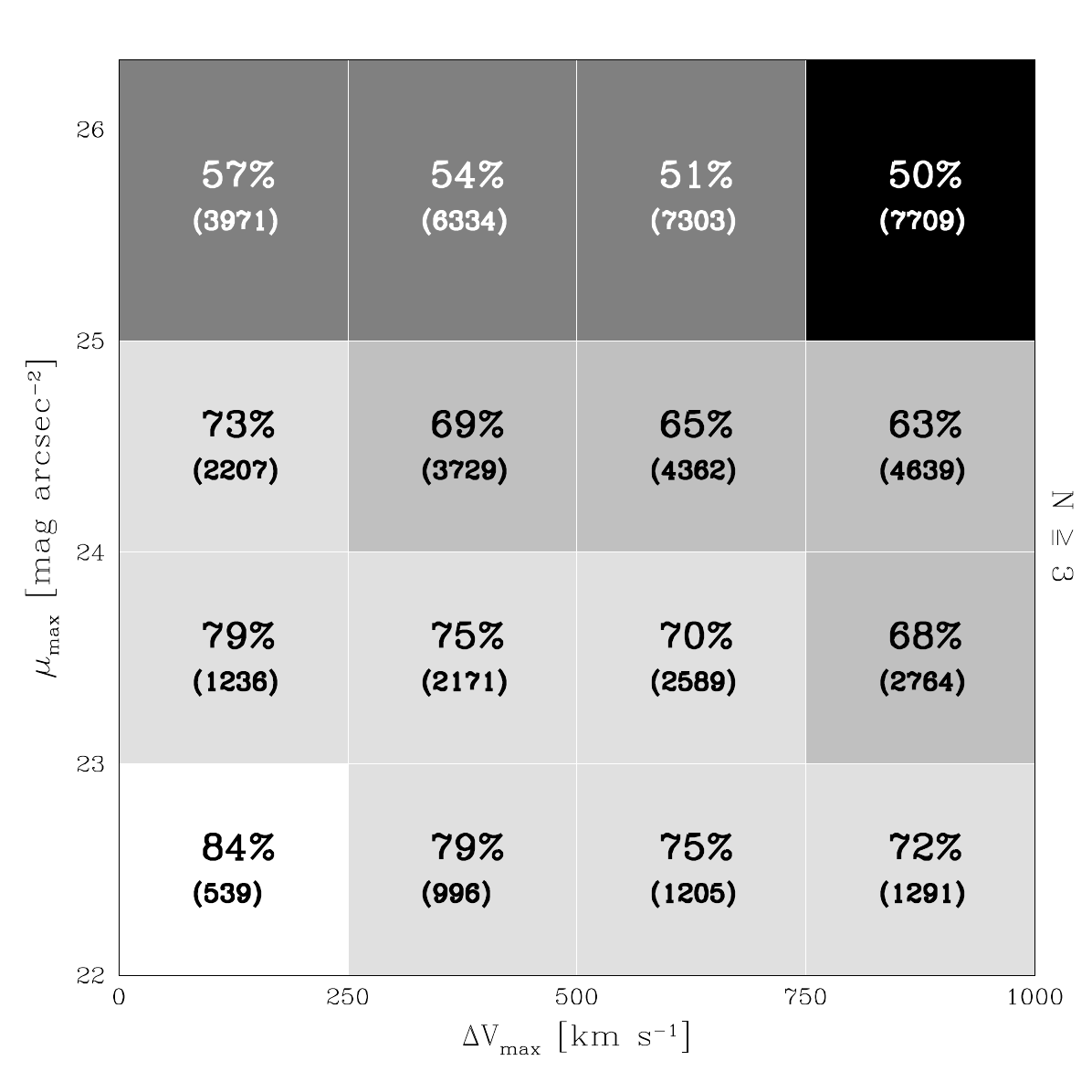}
   \includegraphics[width=0.48\textwidth]{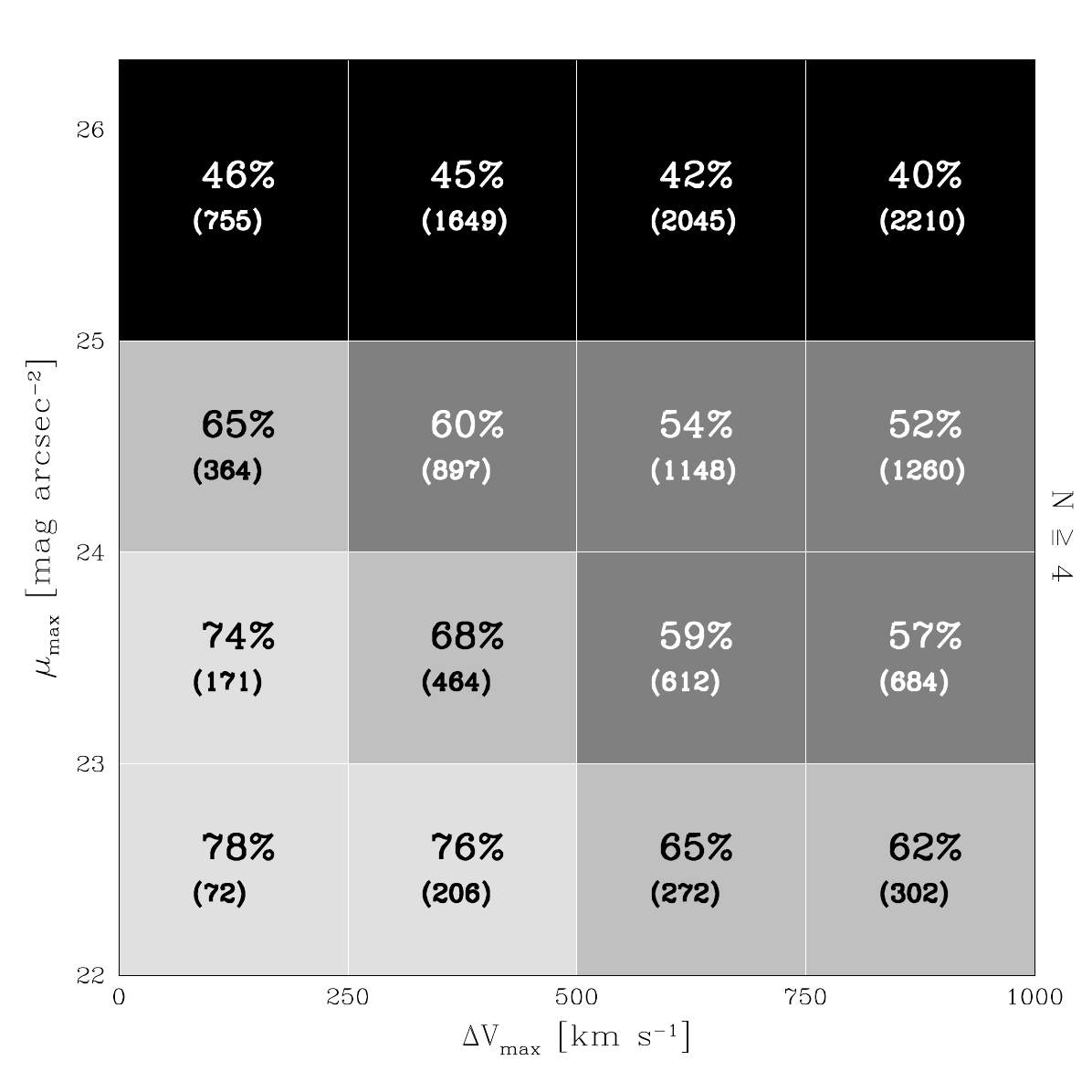}
      \caption{Percentage of robust CGs in $\Delta V_{\rm max}-\mu_{\rm max}$ space that have been observed for over 95\% of the random observers, regardless of their point of view. Each box in this space represents a different CG sub-sample selected with a different $\Delta V_{\rm max}$ and $\mu_{\rm max}$ limits. 
      For instance, the bottom-left box represents the sub-sample with $\mu \leq 23$ and $\Delta V\leq 250$, while the top-right rectangle displays the percentage for the full sample of CGs. Each panel also quotes the total number of CGs (in parentheses) in each resulting sub-sample. Colours are assigned in a five-level greyscale, ranging from the lowest (black) to the highest (white) percentages. The upper plot is performed with CGs with three or more members, while the bottom plot only includes CGs with four or more galaxy members.
      }
         \label{fig:7}
\end{figure}

The top panel of Fig.~\ref{fig:6} clearly shows that only a small percentage (4\%) of CGs fail to meet both criteria, and that lowering either the surface brightness or the velocity-difference maximum of the original CGs leads to a measurable improvement in the fraction of CGs that satisfy both conditions simultaneously.

Finally, because up to this point the parameters $\mu_{\rm max}$ and $\Delta V_{\rm max}$ have been varied one at a time, we now examine which combination of surface brightness and velocity difference maxima applied simultaneously to the original sample maximises the fraction of CGs that satisfy both the compactness and velocity-concordance original criteria, independent of the observer’s position. 
To do this, we explored the parameter space defined by $\Delta V_{\rm max}$ and $\mu_{\rm max}$, selecting pairs of values within this space to restrict the original CG sample and compute the percentage of systems observable by all possible random reference frames according to the original Hickson-like criteria. 
Figure~\ref{fig:7} shows these percentages for each sub-sample defined in this parameter space. For each sub-sample, we also indicate the number of CGs that comprise each sub-sample after applying the corresponding restrictions.
In the upper plot of Fig.~\ref{fig:7} we show the results obtained for those CGs with three or more galaxy members, while the bottom plot only displays the results for those CGs with four or more members (i.e. excluding triplets). 

Focusing on the upper plot, the results show that, starting from the original CG sample (top right corner), one can select a pair of parameters in this space that significantly increased the fraction of robust systems: from 50\% in the original sample to 84\% for a sub-sample with $\mu\le 23 \ {\rm mag \ arcsec^{-2}}$ and $\Delta V \le 250 \ {\rm km \ s^{-1}}$. The cost of this improvement is a substantial reduction in the number of systems included: achieving an increase from 50\% to 84\% reduces the sample to roughly 7\% of the size of the original CG catalogue. 
A slightly less restrictive choice can yield 79\% robust systems by adopting $\mu\le 24 \ {\rm mag \ arcsec^{-2}}$ and $\Delta V \le 250 \ {\rm km \ s^{-1}}$, resulting in a sample about 16\% the size of the original, at least more than twice the sub-sample that maximises the robust CG fraction. Therefore, it is ultimately up to the user of a CG catalogue identified with a Hickson-like criterion to decide which combination of parameters best meets the requirements of a particular statistical study of CGs.

On the other hand, we examined whether these results are related to the membership of CGs, that is, whether there are differences when only systems with four or more members are considered. In the lower panel of Fig.~\ref{fig:7}, we show the effect of excluding triplets from the CG sample. We observe the same trends as those previously described, but the percentages of robust groups are systematically lower than for CGs with at least 3 galaxy members. We performed Barnard's exact test \citep{barnard_test} to disentangle when this statement is statistically correct. We obtain that 13 of 16 sub-samples are statistically consistent (p-value lower than 5\%) with the statement that the percentages obtained excluding triplets are systematically lower. Only for the three sub-samples of the lower left corner, $\mu\le 24$ - $\Delta V \le 250$,  $\mu\le 23$ - $\Delta V \le 250$, and $\mu\le 23$ - $\Delta V \le 500$, we obtain p-values of 9\%, 11\% and 20\%, respectively, that cannot rule out the null hypothesis. This result may indicate that the greater the membership, the more difficult the chance of identifying robust CGs.

\begin{figure}
   \centering
   \includegraphics[width=0.49\textwidth]{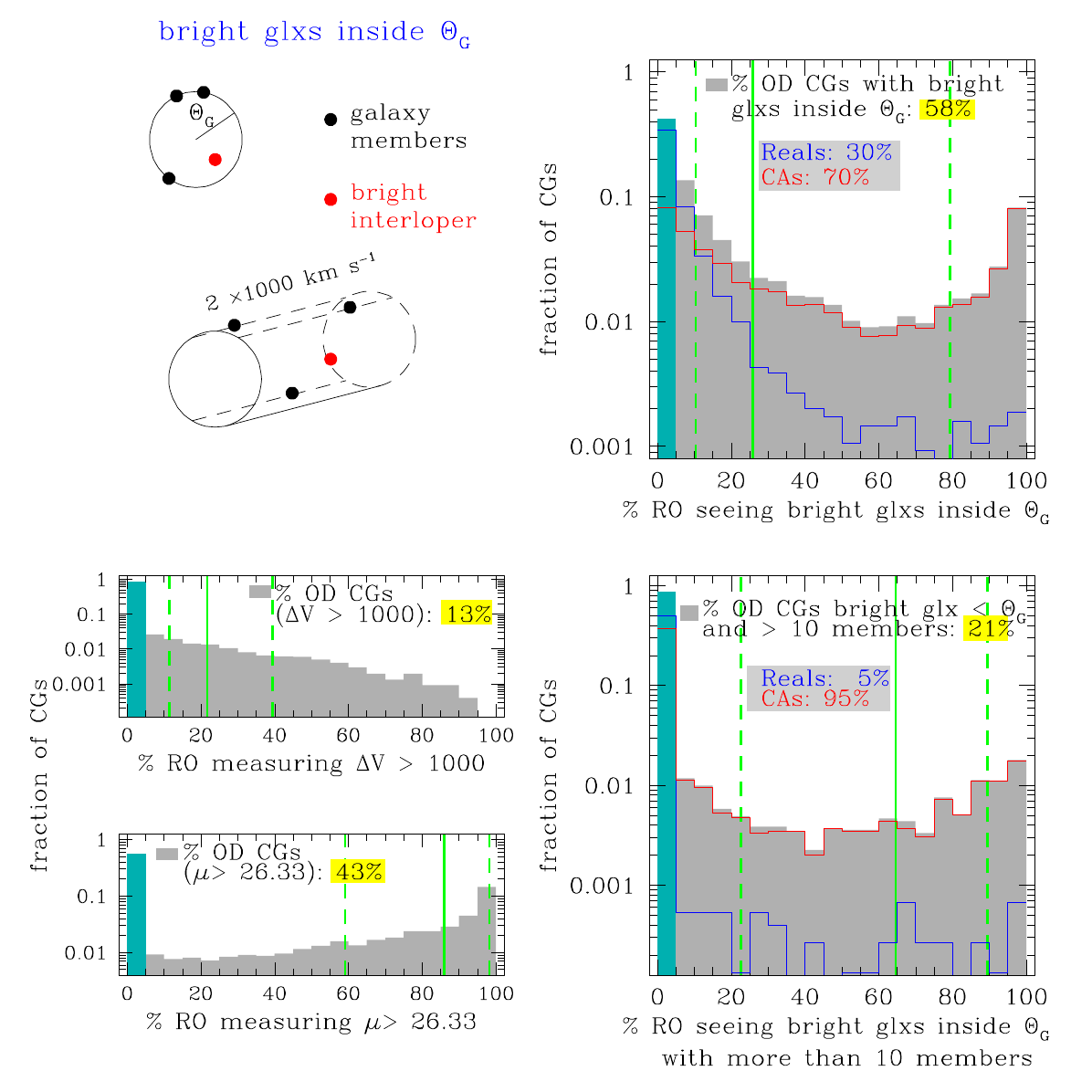}
      \caption{Bright galaxy interlopers inside CGs for different random observers. 
      Top-left panel: Schematic representation of a bright interloper (red) detected by a random observer inside the radius of the CG members (black) and within the $\pm$1000 km/s cylinder. 
      Top-right panel: Distribution among the 7488 CGs of percentages of random observers (RO) who detect bright interlopers.
      The inset legends quote the percentages of observer-dependent (OD) CGs due to the detection of bright interlopers within the system radius. Vertical green lines represent the medians (solid) and the 25th and 75th percentiles (dashed) of the distribution for observer-dependent CGs (grey histograms). The teal bar between 0 and 5\% represents the fraction of CGs considered robust. The blue and red histograms correspond to the distributions for CGs classified as Reals and CAs+Fake, respectively. 
      Bottom-right panel: Same as top-right panel, but only for those CGs with numerous bright interlopers within $\Theta_G$, which leads the number of galaxy members to exceed the maximum (ten) of the population criterion. 
      Bottom-left panels: Same as Fig.~\ref{fig:2} but recalculated for all CGs now including bright interlopers.
      }
         \label{fig:8}
\end{figure}

\subsection{Does the possible detection of bright galaxies not in the original CG influence the detection of the CG from another line of sight?}
Up to this point, we have tested how different observers can detect whether a CG, as seen from the original line of sight, satisfies the Hickson compactness and velocity concordance criteria. To do this, we carried out an experiment involving only the original members of the CGs and analysed how they would be observed from other viewing directions. However, changing the line of sight may not only change the geometry of the system but also lead to the appearance of additional galaxies along that direction, which could affect the detection of the CG.

To assess the presence of these interloper galaxies from other lines of sight, we modified the experiment previously described in subsection~\ref{sec:randomexp}. Instead of using only the CG galaxy members, we selected all galaxies located within a 5~$h^{-1}\, {\rm Mpc}$ sphere around the CG centre in 3D real space. As a result, each random observer examines a considerably large region around the CG location. Nevertheless, some CGs that were previously classified as Fake are so extended in real space that they have galaxy members that do not fall within this sphere. These 221 systems were excluded, leaving a final sample of 7488 systems for this analysis.

Once interlopers are taken into account, both the population and isolation criteria may also be affected. The population criterion requires between three and ten galaxies within three magnitudes of the brightest galaxy; groups with fewer than three or more than ten members are discarded. The isolation criterion, in turn, requires the absence of other bright galaxies within the isolation ring (three times the group size); the presence of any interloper leads to the rejection of the group.

The first scenario corresponds to when a galaxy within a three-magnitude range from the brightest CG galaxy or brighter (hereafter referred to as a bright interloper) lies within the angular radius of the system ($\Theta_{\rm G}$) and within 1000 km $\rm s^{-1}$ of the system centre; therefore, the population criterion is compromised.
In the top-left of Fig.~\ref{fig:8}, we show an example of this situation. The top-right panel shows the distribution among 7488 CG of the percentage of random observers that detect bright interlopers within $\Theta_{\rm G}$ from their perspective. Such interloper galaxies are detected in 58\% of the CGs, while the median fraction of random observers that detect them for a given system is $\simeq$ 26\% (green vertical line). In addition, within 58\%, we distinguish how many of these systems were originally classified as Reals or CAs. Roughly 30\% of these systems were Reals, while the remaining were mostly CAs. 
The distributions of the percentage of observers that detect an interloper, separated into Reals (blue) and CAs (red), are also shown in the figure. Typically, a relatively small percentage of random observers detect bright interlopers inside the Real CGs, whereas for CA CGs, the range of percentages of observers that detect interlopers is quite broad, with a very prominent peak at percentages above 90\%. 
This result is expected, since CA CGs tend to be more elongated along a particular direction in real space, and therefore their projected angular size for a line of sight (different from the original one) could be considerably large, favouring the inclusion of galaxy interlopers within the system. 

Using these bright interlopers together with the original members, we re-estimated the projected centre, the centre along the line of sight and the surface brightness, and then we test whether these new systems still satisfy the compactness and velocity concordance criteria (bottom-left panels of Fig.~\ref{fig:8}). We find that 13\% of the CGs become observer-dependent when the velocity concordance criterion is considered, while 43\% of the CGs become observer-dependent when the compactness criterion is analysed. These percentages are very similar to those obtained when only the original CG members were considered (see Fig.~\ref{fig:2}). We also observe that the distributions obtained for the percentages of random observers are quite similar to those obtained previously in Fig.~\ref{fig:2}.
Therefore, we conclude that the inclusion of bright interlopers does not bias the previously obtained results. 

Nevertheless, we find that the number of bright interlopers is relevant since 21\% of CGs with bright interlopers exceed the ten members usually defined as the maximum in Hickson’s population criterion for a system to be considered a CG. For those systems (bottom-right panel of Fig.~\ref{fig:8}), which are mostly CAs systems, we observe a median of 65\% of the random observers that detect failures in the population criterion. 
Another scenario in which bright interloper galaxies may cause difficulties for the population arises when an interloper is brighter than the brightest galaxy of the original CG. When the brightest galaxy changes for a given random observer, the three-magnitude range defined with respect to the new brightest galaxy may no longer include the minimum of three members required for the system when viewed from a different location.
Therefore, we first computed how many CGs change their brightest galaxy for different random observers. We find that 13\% of CGs change their brightest galaxy for at least 5\% of their random observers, but only 0.5\% of the total sample of CGs fail to find the minimum of three members within a three-magnitude range from its brightest. 
Consequently, we can consider that the impact of this scenario, the failure to satisfy the lower bound of the population criterion, is practically negligible compared to the case in which the number of bright interlopers causes the system to exceed the maximum of ten members imposed by the same criterion.

\begin{figure}
   \centering
   \includegraphics[width=0.49\textwidth]{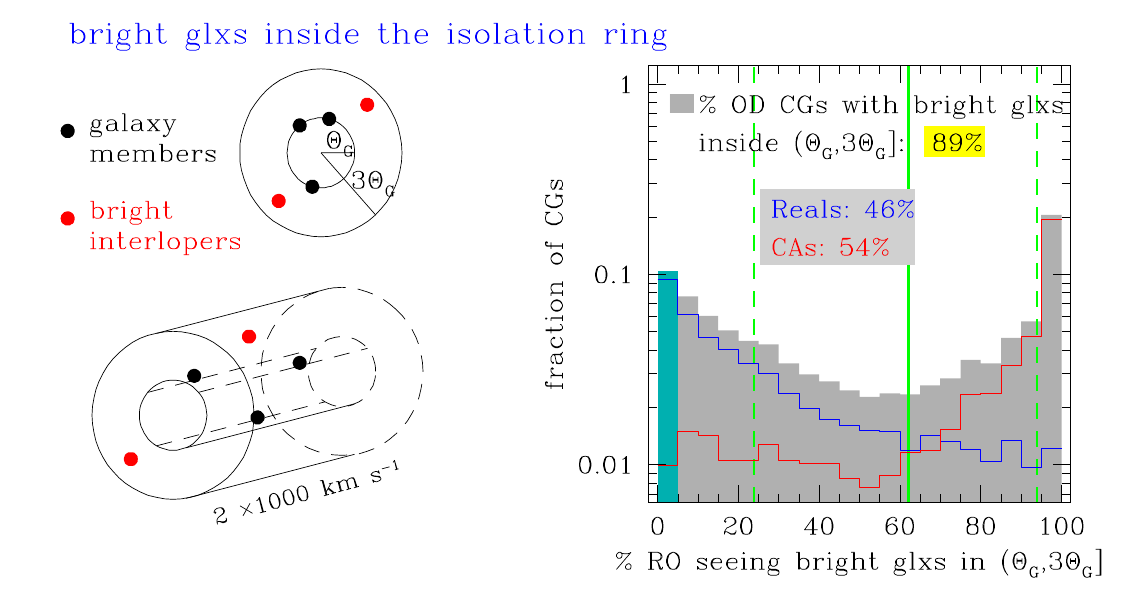}
      \caption{Bright galaxy interlopers inside the isolation ring, ($\Theta_G$,$3\Theta_G$], for different random observers. Left panel: Schematic representation of galaxy members and bright interlopers detected by a random observer within the isolation ring. Right panel: Distribution among the 7488 CGs of the percentages of random observers (RO) who detected interlopers. Inset legends quote the percentages of observer-dependent (OD) CGs due to the detection of bright interlopers in this case.
      Vertical green lines represent the medians (solid) and the 25th and 75th percentiles (dashed) of the distribution for the observer-dependent CGs (grey histograms). The teal bar between 0 and 5\% represents the fraction of CGs considered robust.
      We also display the distributions for CGs classified originally as Reals (blue) and CA+Fake (red) and quote the percentages of these classes of systems with respect to the sample of observer-dependent CGs.
      }
         \label{fig:9}
\end{figure}

The second way in which the presence of a bright interloper can interfere with the detection of a CG is that it will violate the isolation criterion defined by Hickson.  As mentioned earlier, this criterion is not directly related to the apparent compact geometry of the galaxy system but rather intended to select systems with a certain degree of isolation from the surrounding structures. However, the accumulated literature over the years has shown that this criterion is extremely local and that CGs are not isolated systems with respect to the rest of the structures in the Universe. 
\cite{barton+98} were the first to suggest that the vast majority of CGs are embedded in environments ranging from poor groups to rich clusters. In contrast, \cite{decarvalho+05} found that approximately one quarter of their CGs are located within Abell-rich clusters. Similarly, \cite{DiazGimenez+15} reported that only 27\% of the CGs in their catalogue are embedded within loose groups, while \cite{sohn+15} showed that the distribution of the mean local surface density to the 5th nearest galaxy exhibits only a small excess at high densities. However, \cite{mendel+11} reported that about half of their CGs reside within larger groups, a result consistent with \cite{zheng+21} and \cite{taverna+23}, who reported that 50\% and 46\% of their CGs are found in groups and clusters. A more recent study by \cite{tricottet+25} suggests an even higher fraction, with more than 70\% of CGs being embedded in larger systems.
Nevertheless, many CG identifications in different observational catalogues still retain the isolation criterion as part of the general identification scheme proposed by Hickson. Therefore, one may ask whether the application of this isolation criterion by observers with different lines of sight could cause a system to no longer be classified as a CG according to the original Hickson standards. 

To assess the impact of finding a bright interloper within the isolation radius, we used the same experiment described previously for interlopers located inside the system radius, but in this case we simply measured whether there is at least one bright galaxy in the isolation ring (i.e. between one and three times the system radius; see the schematic on the left side of Fig.~\ref{fig:9}). The result of this experiment for the 1000 random observers can be seen in the right panel of Fig.~\ref{fig:9}. We find that $\sim 89\%$ of the CGs become observer-dependent when the isolation criterion is taken into account, with a median of 62\% of random observers detecting such bright interlopers.
When we analyse the observer-dependent CG class, we find that 46\% of these systems can be classified as Reals, while 54\% are CAs. The percentages of random observers who detect interlopers in Real CGs are shifted to lower values, while the opposite is observed in CAs, where all observers find interlopers in these systems. 
Therefore, we conclude that enforcing Hickson’s isolation criterion would affect almost all CGs originally identified. Nonetheless, as noted earlier, this criterion does not strictly fulfil its intended purpose of selecting isolated systems and is not directly related to the compact nature of the galaxy system.

Lastly, we would like to note that these experiments are only a first-order approximation to the problem since they are meant to give us a sense of how the presence of bright interloping galaxies identified by other observers might affect the results.
They are not intended to precisely reproduce the CG identification process as carried out by algorithms that follow Hickson’s criteria. When interlopers are detected, those algorithms run iterative procedures to update group membership and constantly verify compliance with all criteria, something we did not incorporate into our experiments.

\section{Summary and conclusions}
\label{sec:conclusions}

We have investigated how the identification of Hickson-like CGs depends on the observer’s reference frame by using a mock lightcone constructed from the Millennium I Simulation \citep{Springel+05} combined with the semi-analytic galaxies generated by \cite{Guo+11}.
From this lightcone, we built a parent sample of 7709 CGs with standard Hickson-like criteria: population, flux limit of the brightest group galaxy, velocity concordance, compactness, and local isolation. Using the full 3D real space and peculiar velocity information, we were able to `re-observe' every CG from many vantage points and test how often it would still satisfy the velocity concordance and compactness criteria.

Our experiment was designed to mimic the effect of placing 1000 random observers surrounding each group on the surface of a sphere with a radius equal to the group-centric distance in real space, recomputing redshift space positions and apparent magnitudes, and applying the velocity concordance and compactness criteria. Two key results emerged. First, the velocity concordance criterion is only mildly sensitive to observer location: About 10\% of the original CGs fail for at least some observers, and typically only a minority of vantage points lose a given system. 
In contrast, the compactness criterion is far more fragile, as nearly half of the CGs are missed by most observers.
Thus, accomplishing the surface brightness limit is most difficult when different observers intend to recover the original identification performed with a Hickson-like algorithm.

The physical properties of the groups explain these trends. The CGs that remain identifiable from almost every vantage point have galaxy members with relatively low peculiar velocities, while those missed by some observers frequently host galaxies exceeding a velocity modulus of $\sim 1000 \ {\rm km \ s^{-1}}$, making the velocity requirement harder to satisfy along certain lines of sight. For compactness, the decisive factor is the intrinsic 3D shape. Using established real space size metrics, we find that 80\% of the universally recovered systems are `Real’(in the approach defined by \citealt{DiazGimenez&Mamon10}: fairly compact in 3D), whereas 94\% of those lost by most observers are classified as line-of-sight alignments or very extended `Fake’ systems. Although some genuine CGs are occasionally missed (typically those with more than four members whose outer galaxies elongate the projected shape), the dominant effect is the removal of intrinsically elongated structures. In this sense, observer dependence in the compactness criterion acts as a filter that preferentially excludes systems unlikely to be physically dense.

We explored how tightening the Hickson-like limits of an already identified CG catalogue can reduce observer dependence. Lowering the surface brightness threshold from the canonical $23.66 \ {\rm mag \ arcsec^{-2}}$ to $23 \ {\rm mag \ arcsec^{-2}}$ minimises the fraction of compactness-dependent systems from 44\% to about 16\%, though it has little impact on velocity dependence. Conversely, lowering the velocity-difference maximum from 1000 to 250 ${\rm km \ s^{-1}}$ cuts the fraction of velocity-dependent systems from 10\% to below 4\%, with minimal effect on compactness. When both criteria are considered simultaneously, roughly 50\% of the original catalogue is completely observer independent, and failures are dominated by the compactness criterion. Finally, we observed that jointly tightening both parameters produces the largest improvement. For example, selecting original CGs with $\mu\le 23 \ {\rm mag \ arcsec^{-2}}$ and $\Delta V \le 250 \ {\rm km \ s^{-1}}$ yields a sub-sample in which 84\% of groups are observable from any vantage point, though at the cost of shrinking the catalogue to about 7\% of its original size. A more moderate choice of $\mu\le 24 \ {\rm mag \ arcsec^{-2}}$ and $\Delta V \le 250 \ {\rm km \ s^{-1}}$ retains $\sim16\%$ of the systems while still achieving 79\% of robust CGs. 
In addition, we find that when we restrict the total CG sample to systems with four or more members, the fractions of systems that can be observed from any location are systematically lower than those previously reported when triplets are included.

Since our original experiment does not account for bright galaxy interlopers seen from different observers, we analysed how their inclusion could influence our previous results. Our analysis shows that the appearance of bright interlopers when CGs are observed from different lines of sight does not significantly alter the overall robustness of the Hickson compactness and velocity concordance criteria. 
Although bright interlopers are detected in a substantial fraction of systems and may even lead some configurations to exceed the population limit defined by Hickson, the fractions of observer-dependent detections remain very similar to those obtained when only the original group members are considered. 
In contrast, the isolation criterion is highly sensitive to projection effects, and for a large fraction of observers, most groups fail to achieve it. 
Nevertheless, as stated in several previous works, the original sample of CGs is not globally isolated (only local isolation is required). Therefore, contamination by interlopers when observed from other directions is expected. 
Unlike compactness and velocity concordance, population and isolation criteria cannot be tuned after the Hickson-like CG catalogue is created. Both are inherent to the searching procedure in an iterative way. Therefore, there is no easy recipe to minimise the dependence of isolation and population on the observer's point of view. 

It is worth noting that the choice of the semi-analytical model (SAM) adopted in this work may influence the results we present since the fraction of systems classified as physically dense plays an important role in our analysis. Previous studies have shown that the number of CGs expected to be physically dense depends on the chosen SAM. For example, \cite{Taverna+22} showed that the fraction of CGs classified as physically dense (according to \citealt{DiazGimenez&Mamon10}) obtained with the G11 SAM is 1.1, 1.2, 1.1, and 1.8 times higher than that predicted for the \cite{Guo+13,Henriques+15,Henriques+20} and \cite{Ayromlou21} (A21) SAMs, respectively. Except for the A21 SAM, the number of such systems in the other SAMs is similar, suggesting that the results obtained in this work are likely transferable to those models. In the extreme case of A21, where significantly fewer physically dense systems are expected, the results could differ. Nevertheless, a more extensive comparison aimed at disentangling the impact of using different SAMs is beyond the scope of this study.

Our findings lead us to conclude that the dependence of Hickson-like CG identification on the observer’s frame is real (as previously suggested by \citealt{hickson+97}), and it can certainly be quantified, with compactness being the principal source of variation. We observed that velocity-driven failures are limited to systems with high peculiar velocities, which are not as common. In contrast, compactness failures are largely due to intrinsically elongated or chance-alignment configurations. Finally, stricter surface brightness and velocity cuts adopted for the original CGs can dramatically increase the fraction of groups that satisfy both criteria for all possible observers, but such cuts inevitably reduce the number of available systems. Finding the optimal conditions to minimise the detection of systems that are not physically dense is a very interesting tool for overcoming the significant occurrence of such spurious systems in catalogues identified using Hickson-like criteria \citep{McConnachie+08,DiazGimenez&Mamon10}. 

Ultimately, the optimal choice of restrictions to define the CG sample depends on the scientific goals. Studies requiring the most physically compact and statistically robust samples may prefer the more restrictive limits. In contrast, those seeking larger samples for demographic or environmental analyses may tolerate a higher fraction of observer-dependent systems. Our methodology and results provide a quantitative framework for choosing the trade-offs different limits create and highlight the importance of accounting for observer-frame effects in any analysis of compact galaxy groups identified with Hickson-like algorithms. Future work will focus on optimising the searching algorithm to produce a bona fide sample of CGs with lower rates of contamination from scratch.

\begin{acknowledgements}
We warmly thank the referee, Gary Mamon, for his careful reading and his constructive and insightful suggestions that improved the final version of the manuscript.
We thank the authors of the SAM for making their model publicly available. 
The Millennium Simulation databases used in this paper and the web application providing online access to them were constructed as part of the activities of the German Astrophysical Virtual Observatory (GAVO).
This work has been partially supported by the Consejo Nacional de Investigaciones Científicas y Técnicas de la República Argentina (CONICET) and the Secretaría de Ciencia y Tecnología de la Universidad de Córdoba (SeCyT).\\
\end{acknowledgements}

\bibliographystyle{aa} 
\bibliography{refs} 
\end{document}